\documentclass{elsart}

\usepackage{graphicx}

\usepackage{amssymb, amsmath}
\usepackage{longtable}
\begin{document}
\def\expounit{$\rm{km^2\,sr\,yr}$ }
\begin{frontmatter}
\title{On the prospects of  ultra-high energy cosmic rays detection by high altitude antennas}


\author[Phys]{P. Motloch},
\ead{motloch@uchicago.edu}
\author[Astro]{N. Hollon},
\ead{hollon@oddjob.uchicago.edu}
\author[Phys,Astro,EFI,Kavli]{P. Privitera}
\ead{priviter@kicp.uchicago.edu}

\footnotesize
\address[Phys]{Department of Physics, University of Chicago, Chicago, IL, USA 60637}
\address[Astro]{Department of Astronomy and Astrophysics, University of Chicago, Chicago, IL, USA 60637}
\address[EFI]{Enrico Fermi Institute, University of Chicago, Chicago, IL, USA 60637}
\address[Kavli]{Kavli Institute for Cosmological Physics, University of Chicago, Chicago, IL, USA 60637}


\begin{abstract}
Radio emission from Ultra-High Energy Cosmic Rays (UHECR) showers detected after specular reflection off the Antarctic ice surface has been recently demonstrated by the ANITA balloon-borne experiment. An antenna observing a large area of ice or water from a mountaintop, a balloon or a satellite may be competitive with more conventional techniques. We present an estimate of the exposure of a high altitude antenna, which provides insight on the prospects of this technique for UHECR detection. 
We find that a satellite antenna may reach a significantly larger exposure than existing UHECR observatories, but an experimental characterization of the radio reflected signal is required to establish the potential of this approach. A balloon-borne or a mountaintop antenna are found not to be competitive under any circumstances. 
\end{abstract}

\end{frontmatter}

\newpage
\section{Introduction}
The detection of Extensive Air Showers (EAS) by instruments placed at high altitude above ground was proposed as early as 1972 by Chudakov \cite{chudakov}. Cherenkov photons,  emitted in a narrow cone of $\approx 1^{\circ}$ half-angle along the EAS axis, may be diffused after hitting ground (e.g. by snow or water). A Cherenkov detector overlooking the Earth surface from a mountain or a balloon may provide a huge detection aperture at low cost. Several experiments have explored this concept \cite{snow}. Also,  the detection of  UV fluorescence light from Ultra-High Energy Cosmic Rays (UHECR) by a satellite instrument 
is at the core of the JEM-EUSO proposal \cite{jemeuso}.   

ANITA \cite{anita}, a balloon-borne experiment searching for high energy neutrinos through the coherent radio emission from the neutrino-induced shower in the Antarctic ice,  has recently presented evidence of radio detection of EAS \cite{anitaeas}. The characteristics of  the events - polarization and dependence on the geomagnetic field - suggest that the detected radio signal comes from specular reflection off the ice of the EAS highly-beamed geomagnetic radio emission \cite{geosynchr}. A full understanding of the detected signals is still lacking. 
When a data-driven modeling of the measurements is used \cite{anitaeas},  the sample of events is found to have a mean energy of  $1.5\cdot 10^{19}$ eV, and a mean angle of observation relative to the true shower axis of $1.5{^\circ}$. Based on these observations, which suggest that UHECR may be detected with a reasonable aperture by a balloon-borne antenna, a forthcoming ANITA-III flight will include a dedicated trigger for UHECR candidates. 
The proposed EVA mission \cite{eva} - a more sensitive balloon-borne antenna - estimates that several hundreds of events will be detected above $10^{19}$ eV when extrapolating ANITA results to a 50 day flight.  
Recently, a satellite experiment -  the Synoptic Wideband Orbiting Radio Detector (SWORD) - based on the same principle has been proposed~\cite{sword}.  

Since radio signals are minimally attenuated by the atmosphere, a high altitude antenna may detect showers landing at very large distances, potentially providing a large exposure for UHECR. It is thus relevant  to evaluate whether this novel technique can play a role in the next generation of UHECR experiments. In this paper, we have derived the exposure of a high altitude antenna under very general assumptions. While a more accurate estimate requires a detailed knowledge of the radio emission and of the detector system, this study already provides insight on the prospects of this technique for UHECR detection. 

\section{Geometric exposure of a high altitude antenna}
\label{sec:apUHECR}
\begin{figure}[h]
  \begin{center}
    \includegraphics[width=5in]{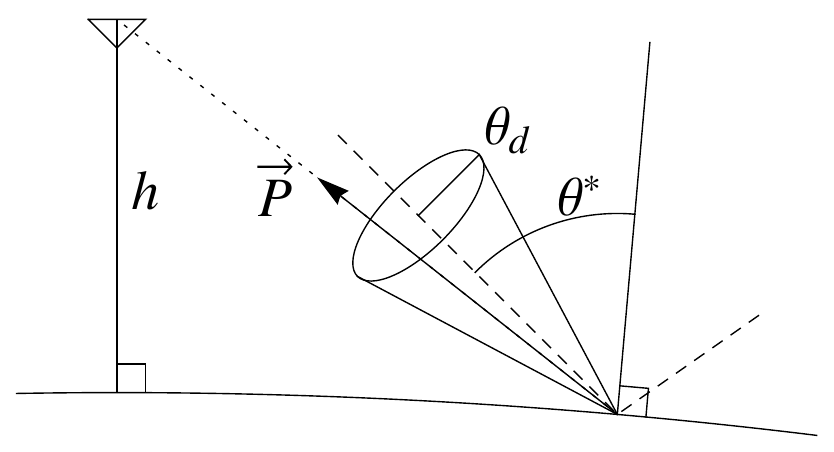}
  \end{center}
  \caption{\small Geometry of cosmic ray detection by a high altitude antenna.}
  \label{fig:geom}
\end{figure}

An analytical estimate of the geometric aperture of a high altitude antenna can be derived under certain approximations. Consider an antenna with an azimuthal field of view of $360^o$ placed at an altitude $h$ above a spherical Earth. To be detected, the specular reflection of the EAS axis is required to be within an angle $\theta_d$ of the direction $\vec{P}$ from the shower impact point to the antenna (see Fig. \ref{fig:geom}).  The antenna is assumed to have 100\% detection efficiency, independent of the shower energy or the distance to the EAS impact point. The geometric aperture is defined as:
 \begin{equation}
A =  \int_S \int_{\Delta \Omega} \cos \theta^{*} d\Omega  dS ,
\label{eq:ap1}
\end{equation}
where $\theta^{*} $ is the angle of the EAS axis with respect to the local zenith at the shower impact point, $\Delta \Omega$ is the detection solid angle of radius $\theta_d$, and $S$ is the area of the spherical cap visible to the antenna. 

The calculation gives:
\begin{equation}
A = 2 \pi^2  \sin^2 \theta_d 
 \frac{  \left[h (2 R_E+h) \right]^{\frac{3}{2}}- h^2 (3 R_E+h)  }{3 (R_E+h)},
\label{eq:ap6}
\end{equation}
where $R_{E}$ is the radius of the Earth. 

\begin{figure}[h]
  \begin{center}
    \includegraphics[width=5in]{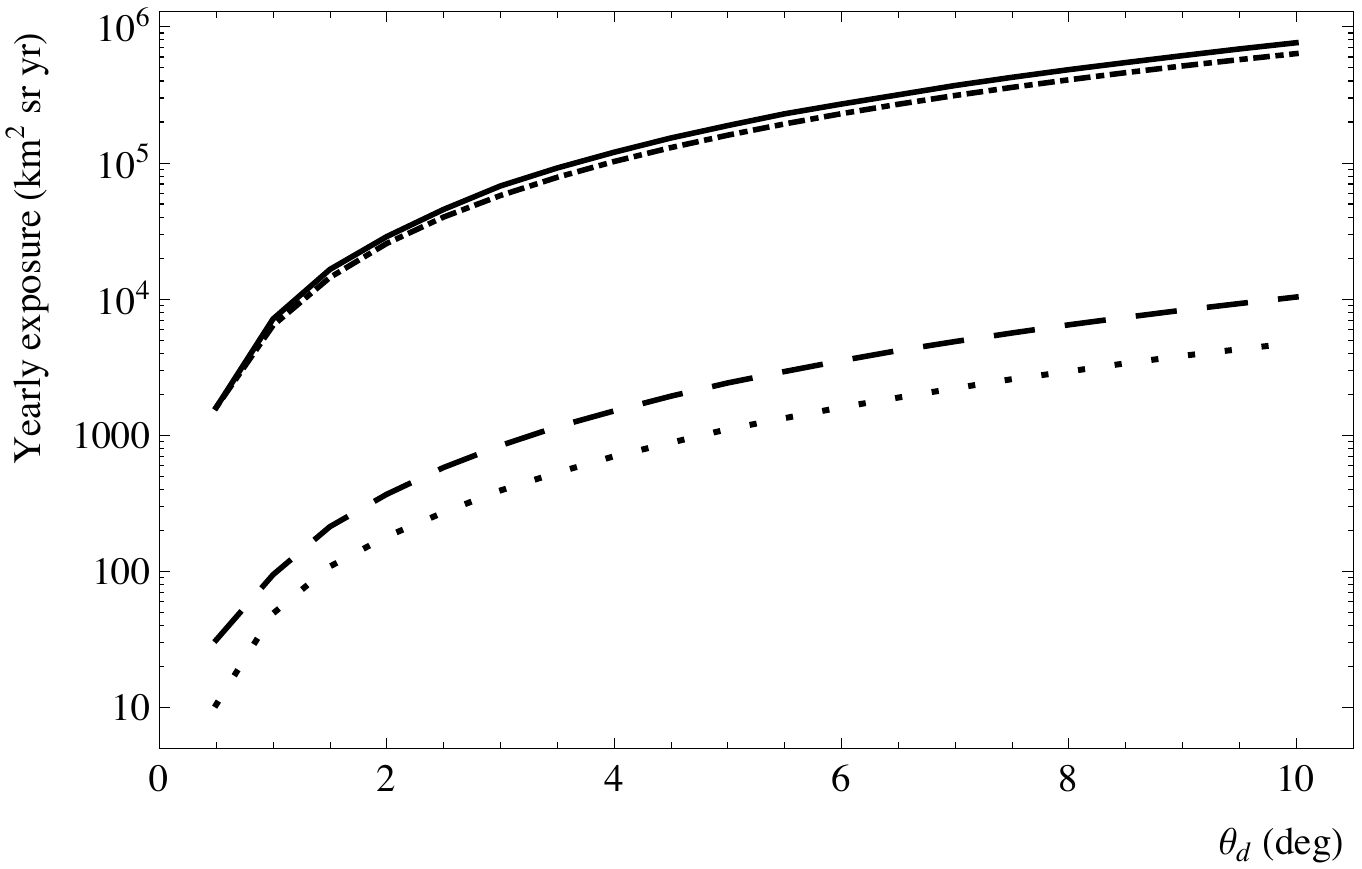}
  \end{center}
  \caption{\small Yearly exposures as a function of detection angle for a mountaintop (dashed line), a balloon-borne (dotted line) and a satellite antenna (solid line). The dash-dot line represents the exposure for a satellite antenna as estimated by an analytical calculation.}
  \label{fig:expo1}
\end{figure}

In deriving Eq. \ref{eq:ap6}, we assumed that the radio emission originates from the shower's impact point at ground. A more realistic estimate may be obtained by taking the position of the maximum development of the shower in the atmosphere as the origin of the radio emission. Since an analytical expression for the geometrical aperture cannot be derived in this case, a Monte Carlo simulation was performed. A uniform distribution of the shower's impact point was generated over the Earth spherical surface. The shower direction was then generated according to an isotropic distribution. We assumed that the radio emission originated at a depth of 850 $\rm{g/cm^{2}}$ (the average shower maximum of a $6\cdot10^{19}$~eV proton \cite{epos199}), and the corresponding point of emission along the shower axis was obtained assuming the US Standard Atmosphere model \cite{usatmo}. The radio emission was parameterized as a cone of half-angle $\theta_d$ around the shower direction, starting at the point of emission. The shower was considered to be detected when the antenna was within the radio emission cone reflected by the spherical Earth surface. 
The time-integrated apertures (i.e. the exposures) derived with this simulation  are given in Fig. \ref{fig:expo1} as a function of the detection angle $\theta_d$ for an antenna located on a  mountaintop, in a balloon, and in a satellite ($h=4,~40,$ and  800~km, respectively). The exposures are calculated for one year of data taking, assuming 13\% duty cycle (50 days flight / year) for the balloon-borne antenna,  and 100\% duty cycle for the other altitudes. 
The exposure for a satellite antenna estimated from  Eq. \ref{eq:ap6} is also given in Fig.~\ref{fig:expo1}. When the altitude of the antenna is much higher than the point of radio emission, the analytical calculation gives a reasonable estimate, and we included  it for reference. 

\begin{figure}[h]
  \begin{center}
    \includegraphics[width=5in]{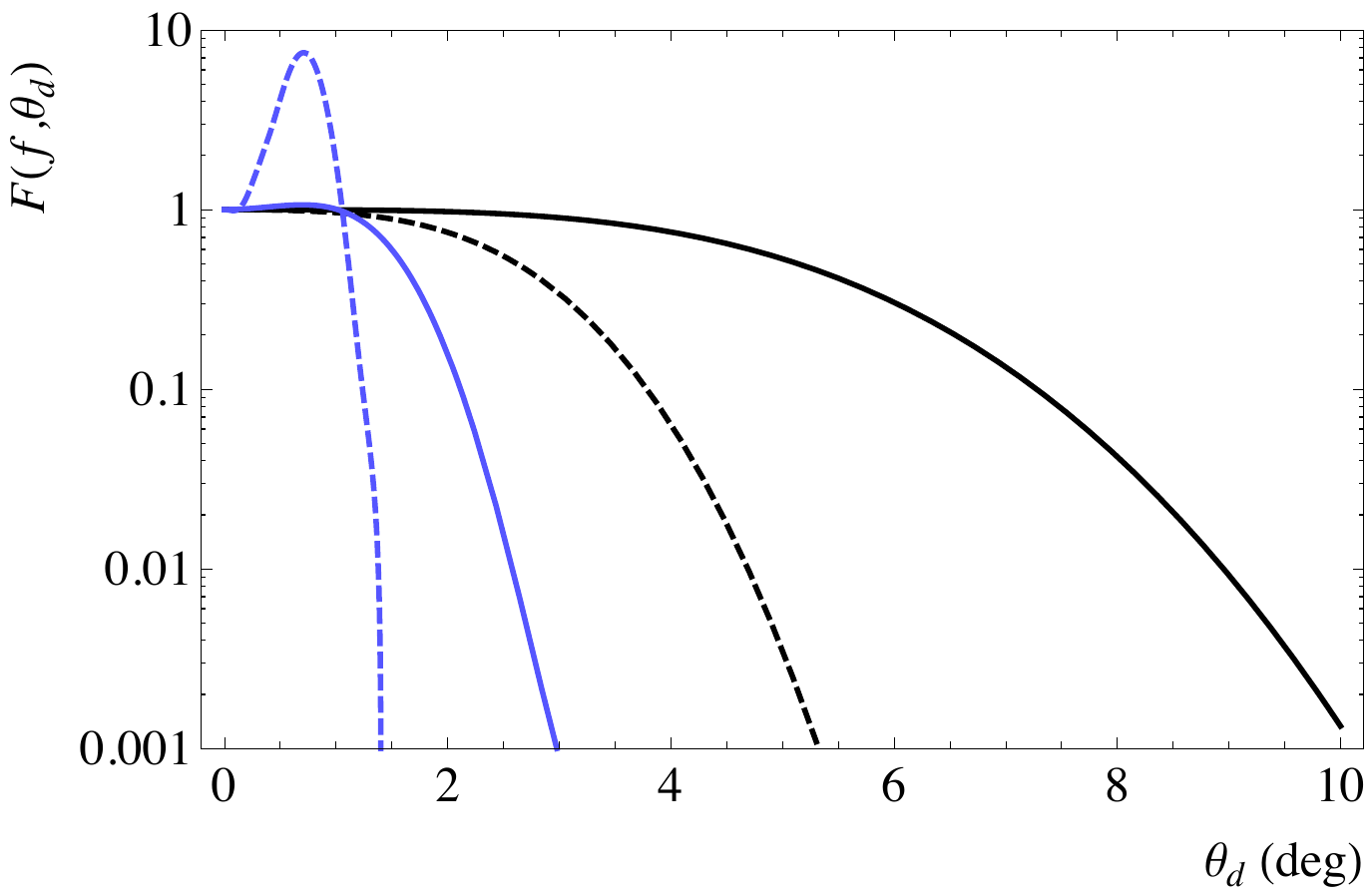}
  \end{center}
  \caption{\small Beam pattern of the cosmic ray radio emission at 30 MHz (solid line) and 200~MHz (dashed line) taken from SWORD (black) and from CoREAS simulations (blue). The curves are normalized to one at $\theta_d=0^o$.}
  \label{fig:beam}
\end{figure}

The actual exposure of an experiment depends on the frequency response of its antenna, since the angular distribution of the EAS radio emission is expected to be frequency dependent. The frequency bands of the ANITA, EVA and SWORD experiments are $f$ = 200-1200~MHz, 150-600~MHz and 30-300~MHz, respectively.  In the following, we will compare two different parameterizations of the beam pattern of the EAS radio emission, $F(f,\theta_{d})$. The first parameterization comes from SWORD \cite{sword}, and is based on the synchrotron radiation formula and ANITA data.  The corresponding beam patterns for frequencies $f = 30$ and 200~MHz are shown in Fig. \ref{fig:beam} (black lines). We obtained a second parameterization with the CoREAS \cite{coreas} simulation package, which is widely used to study radio emission from EAS. Showers of energy $10^{18}$~eV and $5\cdot10^{19}$~eV with zenith angle of $75^o$ \cite{coreassim} were analyzed (a large zenith angle was chosen because most of the aperture comes from distant reflections). The beam pattern was obtained from the electric field at ground assuming the emission point to be at the maximum of the shower development. We found the beam pattern to change minimally with the shower energy. The CoREAS beam patterns are also given in Fig. \ref{fig:beam} (blue lines). The two parameterizations are substantially different, with the SWORD model predicting a much larger beam. Also, the CoREAS beam presents a Cherenkov-like ring pattern which is absent in the SWORD beam model. 

Since the beam pattern becomes narrower for higher frequencies, the maximum aperture is obtained for the lowest frequency of the detection band.  
Let's first consider the SWORD beam pattern. For $f$=150-200~MHz of the ANITA and EVA experiments, the  beam emission drops at  $\theta_{d} \approx 5^o$. The exposure of a balloon-borne antenna for $\theta_d= 5^o$ is $\approx1100$ \expounit  (Fig. \ref{fig:expo1}), which can be taken as an upper limit of the exposure of these experiments. The exposure is somewhat larger for a mountaintop antenna, where the smaller aperture is compensated by the 100\% duty cycle.  For a satellite antenna like SWORD, the lower frequency of the detection band is 30 MHz, which corresponds to a maximum $\theta_{d} \approx 9^o$ (Fig.~\ref{fig:beam}). From Fig.~\ref{fig:expo1}, an upper limit on the exposure of $\approx610000$ \expounit is derived. 

Similar estimates with the narrower CoREAS beam result in only  $\approx 100$ \expounit for a balloon-borne antenna and $\approx 60000$ \expounit for a satellite antenna. This is to be expected, since the exposure depends quadratically on the detection angle ($\sin^2 \theta_d \approx \theta_d^2$ in Eq.~ \ref{eq:ap6}).

Notice that these estimates assume a 100\% detection efficiency, and significantly smaller exposures should be expected for a realistic detector efficiency.
For comparison, the geometric exposure of the Pierre Auger Observatory \cite{augeraperture}  amounts to 7000 \expounit for $\theta^* < 60^o$ and $E > 10^{18.5}$~eV, where the detector is fully efficient. 

\section{Estimate of the exposure with specific models of the radio signal}  
 For a more realistic estimate of the exposure of high altitude antennas, we introduced a parameterization of the EAS radio signal in our Monte Carlo simulation.  We used both a simplified version of SWORD model,  and a parameterization from CoREAS simulations. 
 
\begin{figure}[h]
  \begin{center}
    \includegraphics[width=5in]{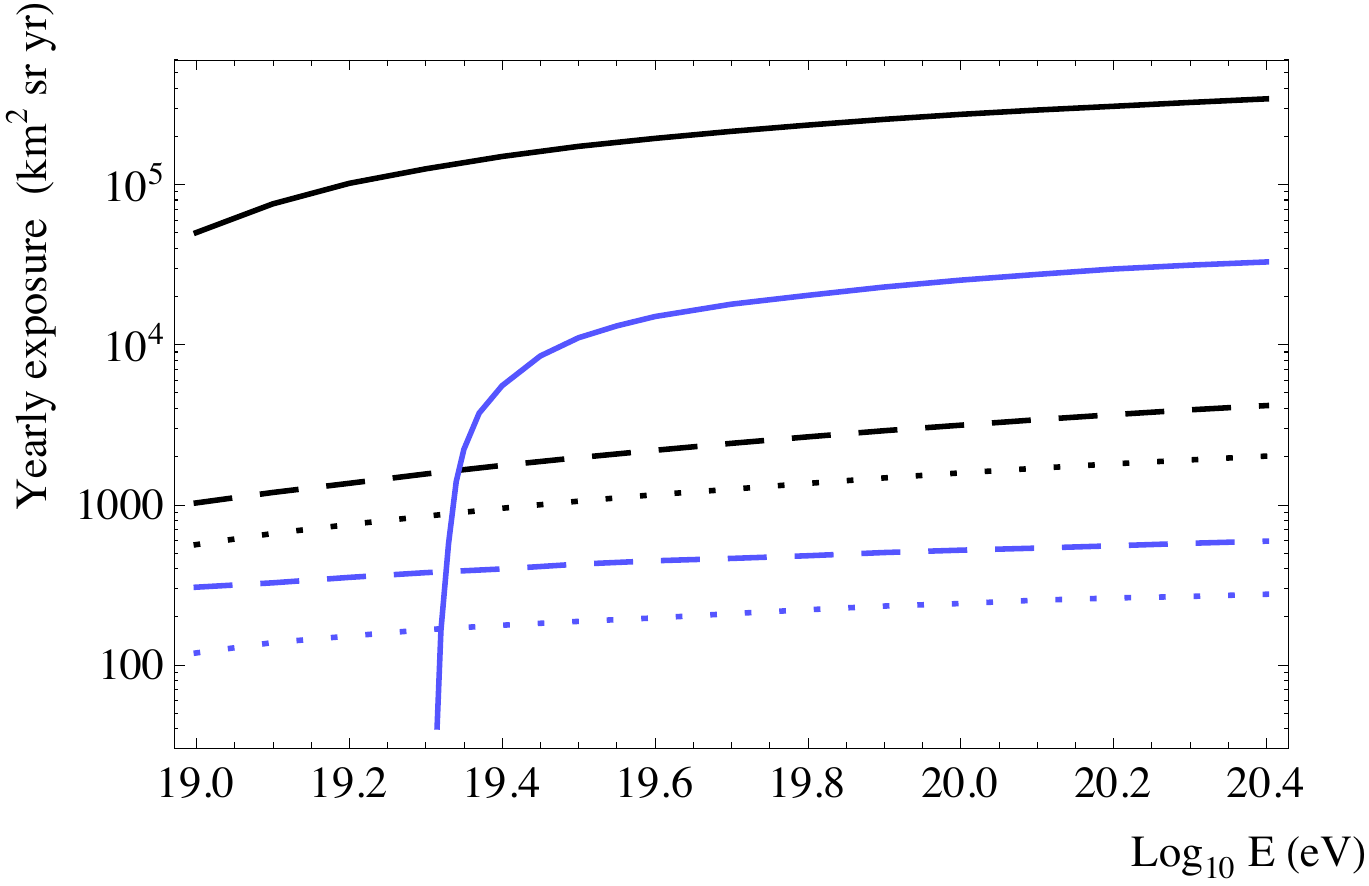}
  \end{center}
  \caption{\small Yearly exposure as a function of energy for a mountaintop (dashed line), a balloon-borne (dotted line) and a satellite antenna (solid line). Radio signal parameterizations based on a SWORD model (black lines) and on CoREAS simulations (blue lines) were used.}
  \label{fig:expdet}
\end{figure}

 The SWORD model was derived from Eq. 1  of \cite{sword}:
 \begin{equation}
 \epsilon(f,E,R,R_{Xmax},\theta_d,\theta^*) =A_0 \left( \frac{E}{10^{19} eV} \right) \left( \frac{R_{ref}}{R+R_{Xmax}} \right) F(f,\theta_d) S(f) \cos\theta^* \mathcal{F}(\theta^*) ,
 \label{eq:signal}
 \end{equation} 
  where $A_0 = 360~\rm{\mu V/m/MHz}$, $R_{ref}=8~\rm{km}$, $R$ is the distance of the antenna to the reflection point on the ground, $R_{Xmax}$ is the distance of the emission point to the reflection point on the ground, $S(f)= \exp \left[ \left( 265~\rm{MHz}- \it{f} \right)/365~\rm{MHz}\right]$ for $f>100~\rm{MHz}$ and $S(f)= \exp \left( 165/365 \right)$ for $f<100~\rm{MHz}$, $F(f,\theta_d)$ is the SWORD beam pattern and $\mathcal{F}(\theta^*)$ is the Fresnel reflection coefficient for electric field polarized parallel to plane of incidence. 
   First, we estimated the exposure of an antenna orbiting at an altitude of 800~km. We simulated showers according to the energy spectrum measured by~\cite{augerspectr}, and we assigned to each shower a signal using Eq. \ref{eq:signal} with $f=45~\rm{MHz}$. At this frequency, the dominant source of noise for a high altitude antenna pointing at the horizon will be the galactic noise background.  For a shower to be triggered, we required its signal to exceed 1.5 times the sky background noise at 45~MHz, estimated by the Cane parameterization \cite{Cane}. 
This simple simulation, which employs a single frequency and detection threshold, reproduces reasonably well the main characteristics - the distributions of zenith angle and detection angle $\theta_d$ - of the events triggered by a satellite antenna as determined in a more sophisticated simulation of the SWORD detector concept given in  \cite{sword}.
Then, we estimated the exposures of a mountaintop and balloon-borne antenna ($h=4$ and 40~km, respectively) using the same signal model and detection threshold of the satellite antenna. These energy dependent exposures are shown in Fig.~\ref{fig:expdet} (black lines). 

We performed an analogous study with a different signal model, which we obtained from the CoREAS simulations:
 \begin{equation}
 \epsilon(f,E,R,R_{Xmax},\theta_d,\theta^*) =A_0 \left( \frac{E}{10^{19} eV} \right) \left( \frac{R_{ref}}{R+R_{Xmax}} \right) F(f,\theta_d) \mathcal{F}(\theta^*) ,
 \label{eq:signalcoreas}
 \end{equation} 
  where $A_0 = 44~\rm{\mu V/m/MHz}$, $R_{ref}=40~\rm{km}$, $f=30$~MHz  and  $F(f,\theta_d)$ is the CoREAS beam pattern (blue solid line of Fig.~\ref{fig:beam}). The corresponding exposures are given in Fig.~\ref{fig:expdet} (blue lines). A trigger threshold 1.5 times the sky background noise at 30~MHz was used. The CoREAS signal model, which may be more realistic since it reproduces reasonably well EAS radio measurements by antenna arrays~\cite{lopes}, predicts significantly smaller exposures than the SWORD model. Notice that Eq.~\ref{eq:signalcoreas} and Eq.~\ref{eq:signal} differ by a factor $\cos\theta^*$, which was empirically introduced in the SWORD model to better fit the ANITA data \cite{romero}. Including this factor in the CoREAS parameterization would further decrease the corresponding exposures in Fig.~\ref{fig:expdet}.   
  
With exposures smaller than few $10^3$ \expounit even in the optimistic case of the SWORD signal model, experiments based on mountaintop or balloon-borne antennas do not look a promising alternative to existing ground arrays. Notice that the effective exposure of the ANITA and EVA experiments, which have a  minimum detection frequency of 150-200~MHz to be compared with 30-45~MHz used in our simulation, should be even smaller than that estimated in Fig.~\ref{fig:expdet} for a balloon-borne antenna.    
Exposures exceeding $\approx 10^4- 10^5$ \expounit could be achieved with a satellite antenna, which makes this experiment  worthwhile further consideration. 

\section{Conclusions}
We have evaluated  the exposure of experiments detecting radio emission from UHECR showers specularly reflected off the ground. We obtained upper limits on the exposures under minimal assumptions on the characteristics of the radio emission. Exposures were also estimated by assuming specific models for the radio emission and detection efficiency. We found that a satellite antenna orbiting at an altitude of 800 km may provide an yearly exposure exceeding $\approx 10^5$ \expounit, more than ten times larger than the exposure of the Pierre Auger Observatory. On the other hand, this prediction is strongly dependent on the modeling of the radio signal, and a significantly smaller exposure - just a few times that of the Auger Observatory - is obtained when state-of-the-art CoREAS radio signal simulations are used.        
A mountaintop or a balloon-borne antenna were found not to be competitive under any circumstances, and  do not appear as a viable alternative for the next generation of UHECR observatories.

While our calculated exposure of a satellite antenna may still look promising, it is likely to be overestimated. In fact,  we assumed the radio reflection efficiency to be the same over the whole Earth surface and we did not take into account the effect of anthropogenic noise. Also, the sensitivity to cosmic ray induced signals is severely affected by dispersion of radio pulses through the ionosphere on their path to the antenna~\cite{sword}, and an efficient trigger scheme has yet to be demonstrated.  
Last, the strength and angular dependence of the EAS radio emission is still quite uncertain. 
Due to these unaccounted effects, the true exposure of a satellite antenna may be significantly smaller than our estimate, which would make this detection approach not competitive with the traditional ground array technique. For a realistic evaluation of the prospects of a satellite based experiment, several measurements need to be performed. A planned pathfinder mission~\cite{chirp} to study the ionospheric dispersion of the radio signal could address the fundamental issue of the trigger. Experiments with a montaintop or a balloon-borne antenna could collect enough events to clarify the radio emission process.  Additional insight is already expected from the forthcoming flight of ANITA-III, which will include a  trigger specifically for UHECR. A dedicated experiment using a mountaintop antenna would collect an even larger sample of UHECR showers, and may be a necessary step to validate a satellite mission.
 
It should be noticed that the energy and angular resolution of UHECR events detected by a high altitude antenna will be worse than that typical of ground arrays or fluorescence detectors, which measure the shower characteristics in much more detail. Given the steepness of the energy spectrum of UHECR, a good energy resolution is needed to avoid spillover from lower energy showers, which would drastically dilute the sensitivity to anisotropy measurements. A realistic estimate of the energy and angular resolution of a high altitude antenna will only be possible after the EAS radio emission is fully characterized experimentally.   
Also, it is unlikely that the radio technique will provide useful information on the composition of UHECR, which would require sensitivity to the position of the maximum development of the EAS in the atmosphere. These limitations suggest caution in the prospects of this novel technique for future UHECR observatories. However, the potential gain in exposure justifies further investigations.  

\section{Acknowledgements}
We thank A. Romero-Wolf for providing detailed information on the SWORD concept and for his careful reading of the paper. We acknowledge stimulating discussions with P. Gorham, D. Saltzberg, A. Vieregg and E. Zas on the radio detection of UHECR showers. We particularly thank T. Huege for providing the CoREAS simulated showers used in this work. This work was supported by the NSF grant PHY-1068696 at the University of Chicago, and the Kavli Institute for Cosmological Physics through grant NSF PHY-1125897 and an endowment from the Kavli Foundation.



\end{document}